\UseRawInputEncoding

\documentclass[aps, pra, superscriptaddress, notitlepage, reprint]{revtex4-1}

\usepackage[english]{babel}
\usepackage{amsmath,amsthm}
\usepackage{amsfonts}
\usepackage[pdfborder={0 0 0}, colorlinks=true, urlcolor=blue, linkcolor=blue, citecolor=blue]{hyperref}
\usepackage{color}
\usepackage{graphicx}
\usepackage{float}
\usepackage{subfigure}
\usepackage{lipsum}
\usepackage{epstopdf}

\begin{document}

\title{Parameter estimation and quantum entanglement in PT symmetrical cavity magnonics system}
\author{Dong  Xie}\email{xiedong@mail.ustc.edu.cn}
\affiliation{College of Science, Guilin University of Aerospace Technology, Guilin, Guangxi 541004, People's Republic of China}

\author{Chunling Xu}
\affiliation{College of Science, Guilin University of Aerospace Technology, Guilin, Guangxi 541004, People's Republic of China}

\author{An Min Wang}
\affiliation{Department of Modern Physics, University of Science and Technology of China, Hefei, Anhui 230026, People's Republic of China}

\begin{abstract}
We investigate the parameter estimation in a magnon-cavity-magnon coupled system. PT symmetrical two magnons system can be formed in the gain magnetic materials by the adiabatic elimination of the cavity field mode. We show that the optimal estimation will not appear at the exceptional point due to that the quantum fluctuations are the strongest at the exceptional point. Moreover, we demonstrate that the measurements at the exceptional point tend to be optimal with the increase of prepared time. And the direct photon detection is the optimal measurement for the initial state in the vacuum input state. For the open PT symmetrical two magnons system, the quantum fluctuations will greatly reduce the degree of entanglement. Finally, we show that a higher estimated magnetic sensitivity can be obtained by measuring the frequency of one magnon in the PT symmetrical two magnons system.
\end{abstract}
\maketitle

\section{Introduction}
For a closed system, Hermitian Hamiltonian can dominate the unitary evolution of the system. The eigenenergies are real for obeying the conservation of energy. In 1998, Bender \textit{et al.}~\cite{lab1} first found that parity-time (PT) symmetrical non-Hermitian system can have real eigenenergies. Then it pushed the theoretical development of complex extension of quantum mechanics. However, for the PT symmetric quantum mechanics theory of closed system, there is still a lack of self-consistent theory~\cite{lab2} and experimental verification. For open system, non-Hermitian Hamiltonian can be effectively used to describe the process of evolution.

PT symmetrical non-Hermitian system has attracted more and more attention~\cite{lab3}. Recent works~\cite{lab4,lab5,lab6,lab7,lab8,lab9,lab9A,lab10A} have showed that exceptional point can enhance the sensitivity.  It is mainly because that at exceptional point, two or more eigenvalues and their corresponding eigenvectors coalesce, leading to a nondiagonalizable Hamiltonian, which generates an Nth-order root law of eigenfrequency splitting.

Recently,  the cavity magnonics system\cite{lab11A,lab12A,lab13A,lab14A,lab15A} has received increasing interest due to that magnons have very high spin density, low damping rate, and high cooperativity with microwave
photons\cite{lab16A}. Moreover, coupled cavity-magnon polaritons can offer an important platform for exploring PT symmetrical non-Hermitian physics.

 In this article, we investigate quantum metrology in a magnon-cavity-magnon coupled system. By the adiabatic elimination of the cavity field mode, we can obtain a two-dimensional PT symmetrical Hamiltonian, which describes the dissipative
magnon-magnon coupling. Using the error propagation formula and quantum Fisher information, we show that the direct photon detection is the optimal measurement for the initial vacuum input state. The optimal estimation will not appear at the exceptional point due to that the quantum fluctuations are the strongest at the exceptional point. What's more, we demonstrate that the measurements at the exceptional point tend to be optimal with the increase of prepared time. For the initial thermal state with different temperatures, we find that the estimation precisions around the exceptional point have little to do with the initial state.  And for the open PT symmetrical system, the quantum fluctuations will greatly reduce the degree of entanglement as opposed to the case of closed PT symmetrical system. Finally, we show that the higher estimated magnetic sensitivity $10^{-19}\mathbf{T} Hz^{-1/2} $ can be achieved for the weak coupling.
It is six orders of magnitude higher than that of the state-of-the-art magnetoelectric sensors.

This article is organized as follows. In Section II, we introduce the physical model of cavity magnonics system and the corresponding PT symmetrical Hamiltonian. In Section III, we obtain the estimation precision of the frequency for one magnon. In Section IV, the degree of entanglement between two magnons is discussed. In Section V, we discuss about the application in measuring the magnetic and analyze  its feasibility in experiment. We make a brief conclusion in Section VI.

\section{model of cavity magnonics system}
We consider that two detuning magnons are coupled to a microwave cavity mode separately. In experiment, on can put two magnetic insulator yttrium iron-garnet (YIG) into a microwave cavity.  Considering small amplitude excitations and using the Holstein Primakoff transformation, the original Hamiltonian of the system can be described  as ($\hbar=1$ throughout this article)
\begin{align}
H=\omega_1a^\dagger a+\omega_2b^\dagger b+\omega_3c^\dagger c+g_{13}(ac^\dagger+a^\dagger c)+g_{23}(bc^\dagger+b^\dagger c),
\label{eq:1}
\end{align}
where $a$, $b$, $c$, $a^\dagger$, $b^\dagger$, $c^\dagger$ are the annihilation and creation operators of the two magnon modes and the cavity field, respectively.
$\omega_1$ and $\omega_2$ are the corresponding resonance frequencies of these two magnon modes. $\omega_3$ is the resonance frequency of the cavity field.
$g_{13}$ and $g_{23}$ represent the coupling strength between the cavity and the magnon modes.

The corresponding quantum Langevin equations including quantum fluctuation are given by
\begin{align}
\dot{a}(t)=-(i\omega_1-\gamma_1)a(t)-ig_{13}c(t)-\sqrt{2\gamma_1}a_{\textmd{\textmd{in}}}^\dagger(t),\label{eq:2}\\
\dot{b}(t)=-(i\omega_2-\gamma_2)b(t)-ig_{23}c(t)-\sqrt{2\gamma_2}b_{\textmd{\textmd{in}}}^\dagger(t),\label{eq:3}\\
\dot{c}(t)=-(i\omega_3+\kappa)c(t)-ig_{13}a(t)-ig_{23}b(t)+\sqrt{2\kappa}c_{\textmd{\textmd{in}}}(t),
\label{eq:4}
\end{align}
where $\gamma_1>0$ and $\gamma_2>0$ denote the gain in magnetic materials~\cite{lab17A,lab18A} and $a_{\textmd{\textmd{in}}}$, $b_{\textmd{\textmd{in}}}$ and $c_{\textmd{\textmd{in}}}$ are the noise operators, which satisfy the following properties
\begin{align}
\langle a_{\textmd{\textmd{in}}}^\dagger(t)a_{\textmd{\textmd{in}}}(t')\rangle=0, \langle a_{\textmd{\textmd{in}}}(t)a_{\textmd{\textmd{in}}}^\dagger(t')\rangle=\delta(t-t'),\\
\langle b_{\textmd{\textmd{in}}}^\dagger(t)b_{\textmd{\textmd{in}}}(t')\rangle=0, \langle b_{\textmd{\textmd{in}}}(t)b_{\textmd{\textmd{in}}}^\dagger(t')\rangle=\delta(t-t'),\\
\langle c_{\textmd{\textmd{in}}}^\dagger(t)c_{\textmd{\textmd{in}}}(t')\rangle=0, \langle c_{\textmd{\textmd{in}}}(t)c_{\textmd{\textmd{in}}}^\dagger(t')\rangle=\delta(t-t').
\end{align}

Considering $\kappa\gg\omega_3, \gamma_1, \gamma_2$, we can perform an adiabatic elimination of the cavity field mode,
\begin{align}
c(t)\simeq\frac{-ig_{13}a(t)-ig_{23}b(t)+\sqrt{2\kappa}c_{\textmd{\textmd{in}}}}{i\omega_3+\kappa}.
\label{eq:4A}
\end{align}

Without loss of generality, we assume that the gain rates of two magnons are equal to each other, i.e., $\gamma_1=\gamma_2=\gamma$, the coupling rates between magnons and cavity are the same, i.e., $g_{13}=g_{23}=g$.

Substituting Eq.~(\ref{eq:4A}) into Eq.~(\ref{eq:2}) and  Eq.~(\ref{eq:3}), we can achieve

\begin{align}
i\frac{d(a(t),\ b(t))^T}{dt}=H_{\textmd{eff}}(a(t),\ b(t))^T-i(A_{in}(t),B_{in}(t))^T,
\label{eq:6a}
\end{align}
where $(A_{in},B_{in})=(\sqrt{2\gamma}a_{in}^\dagger+i\sqrt{2\Gamma}c_{in},\sqrt{2\gamma}b_{in}^\dagger+i\sqrt{2\Gamma}c_{in})$ and the effective Hamiltonian $H_{\textmd{eff}}$ is described as
\[
 {H}_{\textmd{eff}}= \left(
\begin{array}{ll}
\ \omega_1+i(\gamma-\Gamma)\ \ \ \ -i\Gamma\\
-i\Gamma\ \ \   \ \ \ \ \ \ \ \ \ \ \ \ \ \ \ \omega_2+i(\gamma-\Gamma)
  \end{array}
\right ),
\]\label{eq:7}
in which, $\Gamma=\frac{g^2}{\kappa}$.

Let $\gamma=\Gamma$, we obtain the PT symmetrical Hamiltonian as shown in Ref.~\cite{lab10,lab11}
\[
 {H}_{\textmd{PT}}= \left(
\begin{array}{ll}
\ \omega_1\ \ \ \ -i\Gamma\\
-i\Gamma\ \ \   \ \ \ \  \omega_2
  \end{array}
\right ).
\]\label{eq:8}

The eigenvalues of ${H}_{\textmd{PT}}$ are
\begin{align}
E_{1,2}=1/2(\Omega \pm\sqrt{\Delta^2-4\Gamma^2}),
\label{eq:6}
\end{align}
where $\Omega=\omega_1+\omega_2$ and $\Delta=\omega_1-\omega_2$.
When $|\Delta|>|2\Gamma|$, the eigenvalues are real, which presents the PT-exact phase; when $|\Delta|<|2\Gamma|$, the eigenvalues are complex, which presents the PT-broken phase; exceptional point occurs at $|\Delta|=|2\Gamma|$.

The absolute value of the eigenvalues splitting is evaluated as
\begin{align}
\Delta E_{1,2}=|E_{1}-E_{2}|=|\sqrt{\Delta^2-4\Gamma^2}|.
\label{eq:11}
\end{align}
Then, the susceptibility of frequency $\omega_1$ which indicates how sensitive the eigenvalues splitting $\Delta E_{1,2}$ would response to the change of the parameter $\omega_1$ can be expressed as
\begin{align}
\chi_{\omega_1}=\frac{\partial\Delta E_{1,2}}{\partial \omega_1}=|\frac{\Delta}{\sqrt{\Delta^2-4\Gamma^2}}|.
\label{eq:11a}
\end{align}
Based on Eq.~(\ref{eq:11a}), one can see that the susceptibility will approach infinity at the exceptional point $|\Delta|=|2\Gamma|$.
However, it does not mean that the measurement uncertainty of $\omega_1$ will be 0 in its physical condition. Quantum fluctuations~\cite{lab12,lab13} and  statistical noises~\cite{lab14} have shown to be strong at the exceptional point. In addition,
the time required to prepare the system steady-state will diverge at the exceptional point.

\section{estimation precision of the frequency in PT symmetrical system}
In this section, we obtain the estimation precision of the frequency in a finite prepared time. Quantum fluctuations are also included in our results.

The solution of Eq.~(\ref{eq:6a}) can be analytically achieved
\begin{align}
a(t)&=W_{11}a(0)+W_{12}b(0)\nonumber\\
&-\int_0^tds[W_{11}(t-s)A_{in}(s)+W_{12}(t-s)B_{in}(s)],\label{eq:13}\\
b(t)&=W_{21}a(0)+W_{22}b(0)\nonumber\\
&-\int_0^tds[W_{21}(t-s)A_{in}(s)+W_{22}(t-s)B_{in}(s)],\label{eq:14}
\end{align}
in which,
\begin{align}
W_{11}=e^{-i\Omega t/2}[\cosh (\lambda t/2)-\frac{i\Delta}{\lambda}\sinh(\lambda t/2)],\\
W_{22}=e^{-i\Omega t/2}[\cosh (\lambda t/2)+\frac{i\Delta}{\lambda}\sinh(\lambda t/2)],\\
W_{12}=W_{21}=e^{-i\Omega t/2}\frac{-2\Gamma}{\lambda}\sinh(\lambda t/2),
\label{eq:12a}
\end{align}
where $\lambda=\sqrt{4\Gamma^2-\Delta^2}$.

\begin{figure}[h]
\includegraphics[scale=0.8]{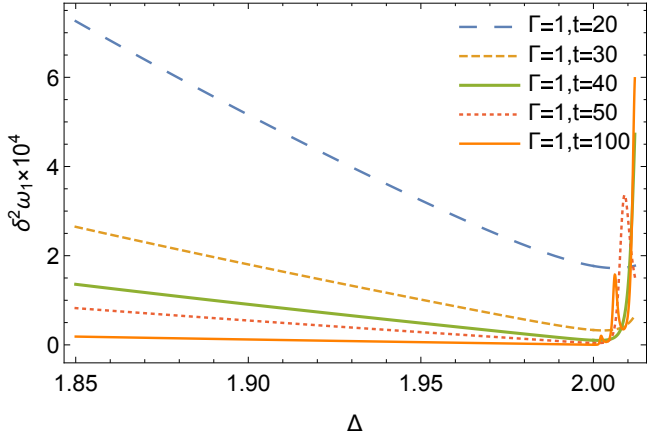}
 \caption{\label{fig.1}Diagram of estimation precision. The estimation precision $\delta^2\omega_1$ changes with the value of $\Delta$ for different evolution time $t$. Here, the value of parameters are given by $\Gamma=1$ and $g/\kappa=0.1$ in the calculations. The exceptional point occurs at $\Delta=2\Gamma=2$.}
 \end{figure}

By directly measuring the number of photon in the cavity, one can obtain the information of $\omega_1$.
The average photon number of cavity reads as
\begin{align}
N_c(t)=\langle c^\dagger(t) c(t)\rangle\simeq\frac{g^2}{\kappa^2}\langle (a^\dagger(t)+b^\dagger(t)) (a(t)+b(t))\rangle.
\label{eq:12a}
\end{align}
Firstly,  we consider that the input state of the two magnon modes is in the vacuum state $|00\rangle$. The corresponding photon number is given by
\begin{align}
N_c(t)\simeq \frac{g^2}{\kappa^2}[\frac{4\Gamma t\Delta^2}{\Lambda^2}-\frac{16\Gamma^3\sin(\Lambda t)}{\Lambda^3}-\frac{8\Gamma^2(1-\cos(\Lambda t))}{\Lambda^2}],
\label{eq:12a}
\end{align}
where $\Lambda=\sqrt{\Delta^2-4\Gamma^2}$.
The variance of $c^\dagger(t) c(t)$ can be described as
\begin{align}
\delta^2N_c(t)=\langle (c^\dagger(t) c(t))^2\rangle-\langle c^\dagger(t) c(t)\rangle^2.
\label{eq:12a}
\end{align}
The following decoupling relation~\cite{lab15} can help to calculate the value of variance
\begin{align}
\langle \hat{A}\hat{B}\hat{C}\hat{D}\rangle\approx&\langle\hat{A}\hat{B}\rangle\langle \hat{C}\hat{D}\rangle+\nonumber\\
&\langle\hat{A}\hat{D}\rangle\langle \hat{B}\hat{C}\rangle+\langle\hat{A}\hat{C}\rangle\langle \hat{B}\hat{D}\rangle-2\langle\hat{A}\rangle\langle\hat{B}\rangle\langle \hat{C}\rangle\langle\hat{D}\rangle.
\label{eq:12a}
\end{align}
For the given initial vacuum state $|00\rangle$, the variance is calculated
\begin{align}
\delta^2N_c(t)&\simeq N_{c}(t)(1+N_{c}(t))
\label{eq:12a}
\end{align}
According to the error propagation formula, the best precision of estimating $\omega_1$ can be evaluated
\begin{align}
\delta^2\omega_1\simeq\frac{\delta^2N_c(t)}{|\partial N_c(t)/\partial\omega_1|^2}.
\label{eq:23a}
\end{align}
The numerical calculation shows that the optimal estimation precision $\delta^2\omega_1$ dose not occur at the exceptional point $\Delta=2\Gamma=2$ as shown in Fig.~1. In addition, we find that the optimal value of $\Delta$ gets closer and closer to the exceptional point with the increase of evolution time $t$. It means that quantum fluctuations prevent the optimal parameter estimation obtained by the given direct photon detection from appearing at the exceptional point. With the increase of prepared time $t$, the effect of quantum fluctuations is suppressed, leading to that the closer the exceptional point is, the higher the estimation precision will be. In other words, the state will be close to the eigenstate of the PT symmetrical Hamiltonian for long time.

Due to that the effective Hamiltonian in Eq.~(\ref{eq:6a}) is a quadratic form, the evolution state will be a Gaussian state~\cite{lab1516} for the given initial vacuum state. Then the quantum Fisher information about the frequency $\omega_1$ can be calculated by~\cite{lab16}
\begin{align}
\mathcal{F}_{\omega_1}=\frac{1}{2}\frac{\textmd{Tr}[(C^{-1}C')^2]}{1+P^2}+2\frac{P'^2}{1-P^4}+(\partial_{\omega_1}\langle\vec{R}\rangle^\mathcal{T})C^{-1}\partial_{\omega_1}\langle\vec{R}\rangle,
\label{eq:24a}
\end{align}
where the first moment is expressed as $\langle\vec{R}\rangle=(\langle{c+c^\dagger}\rangle, \langle{(c-c^\dagger)/i}\rangle)$, the entries of the covariance matrix $C$  is expressed as $C_{ij}=\frac{1}{2}\langle R_iR_j+R_jR_i\rangle-\langle R_i\rangle\langle R_j\rangle$, $P=|C|^{-1/2}$ denotes the purity of the evolution
state, and $X'=\partial X/\partial {\omega_1}$ for $X=C, P$.

After a calculation, for the given initial vacuum state $|00\rangle$, the covariance matrix $C$ reads as
\[
 C=\eta \left(
\begin{array}{ll}
1\ \ \ \ 0\\
0\ \ \   \ 1
  \end{array}
\right ),
\]\label{eq:8}
where $\eta\simeq1+2N_c(t)$. And the first-moment vector is null, $\langle\vec{R}\rangle=0$. Finally, we obtain the quantum Fisher information
\begin{align}
\mathcal{F}_{\omega_1}=\frac{|\partial N_c(t)/\partial\omega_1|^2}{N^2_c(t)+N_c(t)}
\label{eq:24a}
\end{align}
According to the quantum Cram¨¦r-Rao bound~\cite{lab17,lab18,lab19}, the optimal estimation of $\omega_1$ is given by
\begin{align}
\delta^2\omega_1\geq\frac{\delta^2N_c(t)}{|\partial N_c(t)/\partial\omega_1|^2}.
\label{eq:26a}
\end{align}
Comparing Eq.~(\ref{eq:23a}) with Eq.~(\ref{eq:26a}), we find that the direct photon detection can obtain the optimal estimation precision. It means that the direct photon detection is the optimal measurement. The previous result can also be generalized to the case that quantum fluctuations prevent the optimal parameter estimation obtained by the optimal measurement from appearing at the exceptional point.

Then, we consider the initial state as a thermal state, which can be represented by the normalized density matrix
\begin{align}
\rho=(1-e^{-\beta_1})(1-e^{-\beta_2})\exp[-\beta_1a^\dagger a-\beta_2b^\dagger b],
\label{eq:27a}
\end{align}
where $\beta_j=\omega_j/\kappa_BT$ for $j=1,2$ and $T$ represents the temperature.
By the similar calculation, the corresponding average photon number is given by
\begin{align}
&N_c(t)\simeq \frac{g^2}{\kappa^2}[\frac{4\Gamma t\Delta^2}{\Lambda^2}-\frac{16\Gamma^3\sin(\Lambda t)}{\Lambda^3}-\frac{8\Gamma^2(1-\cos(\Lambda t))}{\Lambda^2}+\nonumber\\
&(\frac{t\Delta^2-2\Gamma+2\Gamma\cos(\Lambda t)}{\Lambda^2}-\frac{4\Gamma^2\sin(\Lambda t)}{\Lambda^3})(\frac{1}{e^{\beta_1}-1}+\frac{1}{e^{\beta_2}-1})].
\label{eq:12a}
\end{align}

As shown in Fig.~2, it demonstrates that for the initial thermal state ($T=1$), the optimal estimation precision dose not occur at the exceptional point like the case of the initial vacuum state. From Fig.~3, we find that higher temperature can obtain better estimation precision of $\omega_1$ away from the exceptional point. This is mainly due to the higher number of particles in the thermal state at the higher temperature. However, different temperatures obtain the same estimation precision near the exceptional point. It means that measurements around the exceptional point have little to do with the initial state.

\begin{figure}[h]
\includegraphics[scale=0.8]{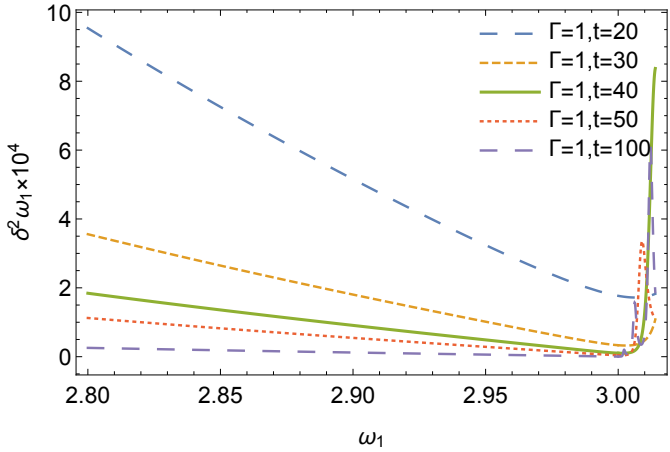}
 \caption{\label{fig.1}Diagram of estimation precision for the initial state in a thermal state. The estimation precision $\delta^2\omega_1$ changes with the value of $\omega_1$ for different evolution time $t$. Here, the value of parameters are given by $\Gamma=1$, $g/\kappa=0.1$, $\omega_2=1$, $T=1$ and $\kappa_B=1$ in the calculations. The exceptional point occurs at $\omega_1=3$.}
 \end{figure}
\begin{figure}[h]
\includegraphics[scale=0.9]{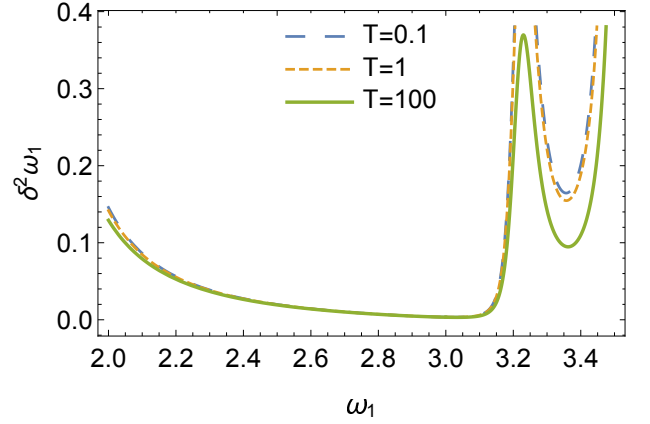}
 \caption{\label{fig.1}Diagram of estimation precision for the initial states in different thermal states. The estimation precision $\delta^2\omega_1$ changes with the value of $\omega_1$ for different temperature $T$. Here, the value of parameters are given by $\Gamma=1$, $g/\kappa=0.1$, $\omega_2=1$, $t=10$ and $\kappa_B=1$ in the calculations. The exceptional point occurs at $\omega_1=3$.}
 \end{figure}
\section{Discussion of Entanglement}
For the Gaussian state, the degree of entanglement between two modes can be effectively quantified by the logarithmic negativity~\cite{lab191}
\begin{align}
E_N=\max[0, -\ln[\nu^{-}]],
\label{eq:29a}
\end{align}
where the smallest symplectic eigenvalue $\nu^{-}=\sqrt{\tilde{\Delta}-[\tilde{\Delta}^2-4\det V ]^{1/2}}/\sqrt{2}$.
$V$ denotes the variance matrix of two magnons, defined by $V_{ij}=\frac{1}{2}\langle U_iU_j+U_jU_i\rangle-\langle U_i\rangle\langle U_j\rangle$. Here, the vector $\mathbf{U}=({a+a^\dagger}, {(a-a^\dagger)/i}, {b+b^\dagger}, {(b-b^\dagger)}/i)$.

The two mode covariance matrix can be described by the general form
\[
 V=\left(
\begin{array}{ll}
\mathbf{A}\ \ \ \mathbf{ C}\\
\mathbf{C}^T\ \ \mathbf{B}
  \end{array}
\right ).
\]\label{eq:8}
The symplectic invariants $\tilde{\Delta}$ can be obtained by $\tilde{\Delta}=\det A+\det B-2\det C$. And $\det(V)$ is the determinant of V.
For the initial state in the vacuum state $|00\rangle$, we can achieve the form of $\nu^{-}$
\begin{align}
\nu^{-}=(\frac{4\Gamma t\Delta^2}{\Lambda^2}-\frac{16\Gamma^3\sin(\Lambda t)}{\Lambda^3}+1)^2-\frac{32\Gamma^4(1-\cos(\Lambda t))^2}{\Lambda^4}].
\end{align}
As shown in Fig.~4, $\nu^{-}$ is still larger than $1$. It means that there is no entanglement between two magnons, $E_N=0$.
\begin{figure}[h]
\includegraphics[scale=0.9]{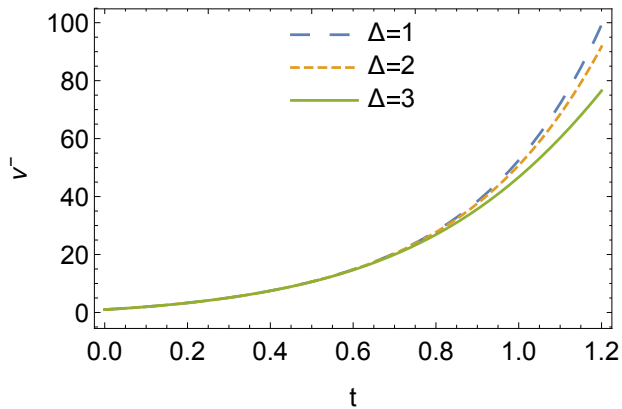}
 \caption{\label{fig.1}Diagram of  the smallest symplectic eigenvalue. $\nu^{-}$ changes with the time $t$ for the initial state as $|00\rangle$. Here, the value of parameters are given by $\Gamma=1$.}
 \end{figure}

For the initial state $|01\rangle$, we check the entanglement between the two magnons by Eq.~(\ref{eq:29a}). We find that there is also no
entanglement. And we adapt the criteria proposed by Hillery and Zubairy~\cite{lab192}. No entanglement has been found. Due to that $|01\rangle$ is non-Gaussian state, these judgement methods are sufficient, but not necessary, for the detection of entanglement. Whether there is entanglement is an open question, which is worth exploring. For the long evolution time, the state will approach the Gaussian state due to the considerably large number of particles and Gaussian noise. As a result, we can obtain that if there is entanglement, it should be small. For the closed PT symmetrical system as considered in Ref.~\cite{lab11}, the pure Bell state (maximal entanglement state) can be obtained. However, for the open PT symmetrical system in this article, we realize that the quantum fluctuations will greatly reduce the degree of entanglement.

\section{Possible application in measuring magnetic field}
For the Kittle mode, the frequency of a magnon linearly depends on the bias magnetic field $B$, which is described as
\begin{align}
\omega_1=\gamma_0B+\omega_{m,0},
\label{eq:27a}
\end{align}
where $\gamma_0 = 28 \mathbf{GHz/T}$ represents the gyromagnetic ratio~\cite{lab20} and $\omega_{m,0}$ is determined by the anisotropy field.
By measuring the parameter $\omega_1$, we can obtain the value of the magnetic field $B$.

Using the feasible parameters~\cite{lab20}, $\Delta=2\pi\times2\textmd{MHZ}$, $\kappa=2\pi\times100\textmd{MHZ}$, and $g_{13}\approx g_{23}\approx 2\pi\times10\textmd{MHZ}$. The gain in magnetic materials can be realized by parametric driving from an ac magnetic field~\cite{lab21}. A recent experiment has demonstrated that a negative magnetic damping  can be induced by the electric field in
heterostructured ferroelectric|ferromagnet layers~\cite{lab22}. To reduce the experimental difficulty for realizing a
negative magnetic damping, Ref.~\cite{lab23} proposed a PT-symmetric synthetic electric circuit coupled with a ferromagnetic sphere.

According to Eq.~(\ref{eq:23a}), we can obtain the estimation precision of the magnetic field $B$ for the evolution time $t=10s$
\begin{align}
\delta B\simeq\frac{\delta N_c(t)}{\gamma_0|\partial N_c(t)/\partial\omega_1|}\simeq 1.8\times10^{-18}\mathbf{T}.
\label{eq:12a}
\end{align}
The corresponding real time magnetic sensitivity defined in Ref.~\cite{lab24} is approximately equal to $10^{-19}\mathbf{T} Hz^{-1/2} $. It is six orders of magnitude higher than that of the state-of-the-art magnetoelectric sensors~\cite{lab24}. The estimated magnetic sensitivity can approach $10^{-15} Hz^{-1/2}$ in the Ref.~\cite{lab23}. But it needs the strong coupling region $g\gg\kappa$. As a contrast, our scheme only requires the week coupling, which is more easily in experiment~\cite{lab20}.
\
\section{Conclusion}
We have investigated the parameter estimation in PT symmetrical cavity magnonics system.  As a result, we show that the optimal estimation will not appear at the exceptional point due to that the quantum fluctuations are the strongest at the exceptional point.
What's more, we demonstrate that the measurements at the exceptional point tend to be optimal with the increase of prepared time.
And the direct photon detection is the optimal measurement.
For the initial thermal state with different temperatures, we obtain the same estimation precisions near the exceptional point. It means that measurements around the exceptional point have little to do with the initial state. Then, we discuss about the entanglement between two magnons and find that there is no entanglement for the initial state in the vacuum state. And for the open PT symmetrical system, the quantum fluctuations will greatly reduce the degree of entanglement. By measuring the frequency, one can obtain the information of magnetic field. Using the feasible parameters, we show that the higher estimated magnetic sensitivity $10^{-19}\mathbf{T} Hz^{-1/2} $ can be achieved for the weak coupling.
Our results will pave the way of the preparation of highly sensitive magnetometers.

\section*{Acknowledgements}

This research was supported by the National Natural Science Foundation of China under Grant No. 62001134 and Guangxi Natural Science Foundation under Grant No. 2020GXNSFAA159047 and National Key R\&D Program of China under Grant No.2018YFB1601402-2.

\end{document}